\documentclass[prd,nofootinbib,floats,superscriptaddress,eqsecnum,tightenlines,11pt]{revtex4}

\usepackage{hyperref}
\usepackage{graphicx}
\usepackage{amsmath,amssymb,amsfonts,amsthm,latexsym,stmaryrd,bbm}
\usepackage{marginnote}
\usepackage[normalem]{ulem}
\usepackage{color}

\def\be{\begin{equation}}
\def\ee{\end{equation}}
\def\bea{\begin{eqnarray}}
\def\eea{\end{eqnarray}}
\def\nn{\nonumber}
\def\SU{\text{SU}}
\def\SL{\text{SL}}
\newcommand{\cG}{{\mathcal G}}
\newcommand{\cL}{{\mathcal L}}
\newcommand{\cH}{{\mathcal H}}
\newcommand{\cM}{{\mathcal M}}
\newcommand{\cD}{{\mathcal D}}

\begin{document}

\title{Spherically symmetric sector of self dual Ashtekar gravity coupled to matter: \\
Anomaly-free algebra of constraints with holonomy corrections}

\author{{\bf Jibril Ben Achour}}\email{jibrilbenachour@gmail.com}
\affiliation{Center for Field Theory and Particle Physics, Fudan University, 200433 Shanghai, China}

\author{{\bf  Suddhasattwa Brahma}}\email{suddhasattwa.brahma@gmail.com}
\affiliation{Center for Field Theory and Particle Physics, Fudan University, 200433 Shanghai, China}

\author{{\bf Antonino Marcian\`o}}\email{marciano@fudan.edu.cn}
\affiliation{Center for Field Theory and Particle Physics, Fudan University, 200433 Shanghai, China}

\begin{abstract}
\noindent
Using self dual Ashtekar variables, we investigate (at the effective level) the spherically symmetry reduced model of loop quantum gravity, both in vacuum and when coupled to a scalar field. Within the real Ashtekar-Barbero formulation, the system \textit{scalar field coupled to spherically symmetric gravity} is known to possess a non closed (quantum) algebra of constraints once holonomy corrections are introduced, which forbids the loop quantization of the model. Moreover, the vacuum case, while not anomalous, introduces modifications which are usually interpreted as  an effective signature change of the metric in the deep quantum region. We show in this paper that both those complications disappear when working with self dual Ashtekar variables, both in the vacuum case and in the case of gravity minimally coupled to a scalar field. In this framework, the algebra of the holonomy corrected constraints is anomaly free and reproduces the classical hypersurface deformation algebra without any deformations. A possible path towards quantization of this model is briefly discussed.
\end{abstract}

\maketitle

\section{Introduction}
\noindent
In the last decade symmetry reduced models have played an increasing pivotal role in Loop Quantum Gravity (LQG). The full theory being not yet accessible, an important effort has been devoted to applying the polymer quantization procedure to some restricted sector of the phase space of General Relativity (GR). The most exciting sectors to apply the loop quantization are the ones which admit classical singularities, typically early universe cosmology and black hole geometries. Indeed, it is expected that taking into account the quantum nature of geometry, through loop quantization, will lead to regular quantum geometries where the singularities shall be naturally resolved. While regular quantum cosmological geometries provide a very promising framework to study bouncing quantum cosmologies and extend the cosmological scenario to the Planck era \cite{LQC0, LQC1, LQC2, Ash2, LQC3, LQC4, LQC5, LQC6, LQC7}, regular quantum dynamical black hole geometries would be the ideal platform to test ideas about the gravitational collapse scenario, Hawking radiation and the tunneling from black holes to white holes recently proposed in \cite{R1, R2, Perez1, BlacktoWhite1, BlacktoWhite2, BlacktoWhite3}.

Let us briefly summarize the strategy used in such (effective) loop models. Once the symmetry reduced phase space is obtained, one follows the loop procedure and polymerize the (components of the) connection variable which survived the symmetry reduction. Physically, this is justified by the fact that, at the quantum level, the gravitational field is not well described by the connection but rather by its holonomy, which is an extended unidimensional object. Then one obtains a new phase space where the symmetry reduced constraints are modified by the so called holonomy corrections. These quantum corrections encode the effect of the quantum nature of the geometry at the Planck scale. Using this strategy and loop techniques, one then quantizes this effective phase space which is believed to be the correct physical one suitable to describe quantum gravity effects.

However, the holonomy corrections introduced in the classical constraints have important consequences on the fate of the symmetries of the system described by the modified gauge generators. Indeed, the first class constraints of GR, i.e. the vectorial constraint $\cH_a$ and the scalar constraint $\cH$, generate the infinitesimal four dimensional diffeomorphisms and form therefore a closed algebra called the hypersurface deformation algebra. It is therefore natural to wonder what is the generalization of the underlying symmetry of the effective phase space with holonomy corrections. Putting it differently, does the algebra of the constraints remain closed after implementing the holonomy corrections? Such questions refer to the covariance of the effective phase space and are therefore of primary importance. Indeed, if the algebra of the modified constraints does not close, it is then impossible to canonically quantize the system following the Dirac procedure since then we are violating the underlying gauge symmetry of the system, which in the case of gravity, is general covariance. One then has lesser number of first class constraints left to generate gauge transformations and we are left with spurious degrees of freedom in the modified system.

Following \cite{S1}, we will say that a system admits a \textit{covariant} quantization if the following two requirements are satisfied:
\begin{enumerate}
\item The first class constraints remain first class after implementing the holonomy corrections. They correspond therefore to infinitesimal generators of some symmetries for the phase space variables and form a closed algebra (at the effective level).
\item The modified first class constraints admit the right classical limit and lead to the usual hypersurface deformation algebra of GR in that limit.
\end{enumerate}

Up to now, the most important effort in the loop quantization of symmetry reduced models has been spared while focusing on the homogenous  case, which have grown into the sub-field of Loop Quantum Cosmology (LQC) \cite{LQC0, LQC1, LQC2}. From those models, one can exhibit some generic results inherited by the polymer quantization -- the most important one being the resolution of the initial cosmological singularity. This singularity resolution allows  one to extend the well known standard model of cosmology, starting at the beginning of inflation, to the so called Planck era and even beyond \cite{Ash2, LQC3, LQC4}. This extension to the pre-inflationary era led to a very large literature, and it is now argued that the LQC framework is now mature enough to make contact with the observations \cite{LQC5, LQC6, LQC7}.

Yet, the symmetry reduction implemented within the context of LQC is a very drastic one, where only the Hamiltonian constraint survives in a very simplified form. In this over-simplified framework, the covariance of the holonomy corrected system becomes trivial due to vanishing of the spatial diffeomorphism constraint. However, the question of covariance can become highly non trivial when applied to other symmetry reduced models, as pointed out in \cite{S1, S2}. In order to investigate this property within the holonomy corrected loop models, one needs to go beyond the homogenous  framework, and turn to less drastic symmetry reduced models where, at least, some components of the Gauss constraint $\cG_i$, the vectorial constraint $\cH_a$ and the scalar constraint $\cH$ survive. Only in this context can one investigate the fate of the covariance of the effective system by studying the closure of the algebra of  modified constraints, and study possible deformations to it. This task was worked out for the spherical symmetric case in \cite{S1} and for the Gowdy model in \cite{S2}, using real Ashtekar-Barbero variables.

Let us summarize those results. In the spherically symmetric case, one needs to distinguish between the vacuum case, which does not have local degrees of freedom, and the system where matter is coupled to gravity, which exhibits local physical degrees of freedom. Indeed, for the vacuum case, the algebra of the modified constraints remain closed although some modifications show up in the structure functions appearing in the Poisson bracket of the scalar constraint with itself $\{ \cH[N_1], \cH[N_2] \} $. This deformation in the algebra has been interpreted in the literature as an `effective' change of signature of the metric in the deep quantum regime, but the physical nature of this modification is not yet fully understood at the fundamental level. Such phenomenon leads to drastic modifications such as the loss of metric structure and a change in the character of the partial differential equations involved from hyperbolic to elliptic \cite{B0, B1, AJ}. See e.g. \cite{B1bis, Miel1, Miel2, Miel3} for some investigations of the consequences in LQC, and \cite{Miel4} for the resulting deformations in the Poincar\'e algebra in the flat spacetime limit. Yet, it is also possible that such phenomenon is a pure mathematical artifact due to the way non-perturbative quantum corrections are implemented. We note that this signature change shows up also in the so called anomaly free approach to compute the cosmological perturbations as developed in \cite{G1, G2}. See \cite{B2} for a detailed discussion. Despite this point, ignoring the signature change phenomenon, the spherically symmetric vacuum can be safely quantized using loop techniques. However, the situation gets quite messy when one tries to couple some local matter degrees of freedom. Indeed, when coupling the spherically symmetric gravitational field to matter, the holonomy modification function prevents the Poisson brackets of the scalar constraint with itself from closing into the diffeomorphism constraint. Thus covariance is violated in the case of the modified system \textit{gravity plus matter}. Therefore, already for the spherically symmetric case, this partial no go result prevents from coupling any kind of matter to gravity in the loop approach. The situation is much worse in the context of the Gowdy model where the same no go result show up already at the level of the vacuum case (which contains gravitational local degrees of freedom) \cite{S2}. Therefore, these powerful no go results derived in \cite{S1, S2} represent a general obstruction to the development of a loop quantization of midisuperspace models that have some local physical degrees of freedom.

Having clarified the situation from the point of view of the underlying covariance of the model, let us now summarize what has been done in the loop quantization of the spherically symmetry reduced model. Since the vacuum spherical case remains covariant even after implementing the holonomy corrections, one can safely complete the quantization. The polymer quantization of the Schwarzschild interior spacetime was worked out first by Ashtekar and Bojowald almost ten years ago \cite{Ash1}. Their quantization is based on the fact that the interior Schwarzschild spacetime can be described as a homogenous contracting cosmology, i.e. the Kantowski-Sachs homogenous space-time. This quantization was studied further by Modesto in \cite{LM1, LM2}. Following the improved dynamics introduced in LQC, Bohmer and Vandersloot introduced a new scheme in \cite{VD1, VD2}, fixing some drawbacks of the previous works, such as the dependency on some auxiliary unphysical structures as the fiducial cell. The interior problem was then revisited in \cite{C1} by Campiglia, Gambini and Pullin. The inhomogeneous exterior space-time was worked out in \cite{C2} by the same authors, some of whom extended the study to the whole space-time in \cite{C3}. In these works, the diffeomorphism constraint was removed by a suitable gauge fixing. In \cite{C4}, Gambini and Pullin introduced a new quantization procedure, based on an Abelianization of the constraints, avoiding therefore the previous gauge fixing of the diffeomorphism constraint. More recently, the interior problem was revisited by Corichi and Singh in \cite{CS1}, improving the classical limit of the antecedent works aforementioned. As expected, a common conclusion of these works is that the central singularity is removed by the polymer quantization procedure, leading to a vacuum regular quantum geometry for the black hole. Unfortunately, as mentioned earlier, the loop quantization of the full dynamical system gravity plus matter remains elusive up to now, because of the difficulty of closing the modified algebra. An exception of this statement is the electro-vacuum case which was treated in \cite{GJM}. Because of these difficulties, the first quantization of a scalar field coupled to spherical symmetric gravity were initiated in \cite{J1, J2} using another strategy called uniform discretization technique \cite{UD}. In \cite{J3, J4}, a new strategy was proposed based on the new gauge fixing leading to a simplification of the modified constraints. Although interesting, such gauge fixings prevent us from inferring anything about the covariance of the system. Few years later, Gambini and Pullin introduced the first study of the Hawking radiation in this framework \cite{J6, J7}, while the quantization of a test shell was presented in \cite{GJ1}. While very promising, it is important to note that those conclusions rely on the study of a test scalar field over a quantum spherically symmetric vacuum geometry and thus not on the full quantization of a scalar field coupled to spherical symmetric gravity.  These models typically have a non-matching version of covariance of the matter and the gravity sectors, as explained in \cite{S1}. Although those results on the vacuum black hole geometry are very promising and lead to interesting insights, it seems mandatory to go beyond the vacuum case and obtain a quantizable model for the full system \textit{scalar field coupled to spherically symmetric gravity}. In order to so, one needs to by pass the partial no go results of \cite{S1} associated with the algebra of the modified constraints for this system.

The first attempt to go beyond the test field approximation and study a full spherically self gravitating collapsing shell was introduced recently in \cite{M1}. In this model, the authors succeeded to bypass the no go theorem of \cite{S1} and obtained an anomaly free algebra of the modified constraint. The strategy used in this work is to keep the Poisson bracket unchanged so that $\{ K_{\phi}, E^{\phi}\} = \{ f(K_{\phi}),  g(K_{\phi}) E^{\phi}\}$ which is then a canonical transformation for a suitable choice of $f$ and $g$, namely $f(K_{\phi}) = \sin(\rho K_{\phi}) / \rho$ and $g(K_{\phi}) = 1 / \cos(\rho K_{\phi})$. While this canonical transformation provides indeed an anomaly free algebra of constraints, there are three difficulties arising within this approach. First, because we are dealing with a canonical transformation of the classical case, it is then difficult to interpret them as loop quantized models where, usually, the effective Poisson bracket gets modified due to polymerization. Secondly, while the first modification function $f(K_{\phi})$ accounts for the usual polymerization of the connection component, the second modification function $g(K_{\phi})$ remains quite unusual with respect to the loop quantization procedure. Indeed, it is not clear why one should implement a holonomy correction on the conjugate triad variable $E^{\phi}$ in the process and this step remains to be justified from first principles. Finally, it is immediate to see that at the singularity, where $f(K_{\phi}) = \sin(\rho K_{\phi}) / \rho$ becomes maximal, the correction $g(K_{\phi}) = 1 / \cos(\rho K_{\phi})$ is no more well defined. One, therefore, has to remove by hand some regions from the spectrum of the theory to obtain a well defined effective theory, which seems problematic if one wants to describe the bounce at the dynamical level. While interesting for evading the no go result of \cite{S1}, this proposal turns out to suffers from drawbacks that still need to be fixed or clarified, either through the quantization of the model, or through some new inputs. It is therefore natural to look for another perspective in order to answer the question of the anomaly freeness of this system.

In this paper, we introduced a different strategy to obtain an anomaly-free algebra of constraints which does not suffers from the same drawbacks. The initial observation is to wonder if the choice of variables is responsible for the anomaly of the modified algebra. Indeed, all the conclusion and partial no go results presented in \cite{S1,S2} were obtained using the real Ashtekar-Barbero formulation. How many of those conclusions might get modified if one uses instead the self dual variables? (Interestingly, a similar comparison between the Schr\"odinger quantization of spherically symmetric models, using real variables and the self-dual variables, was done in \cite{WDWSS}).

Recently, some efforts have been developed to understand more precisely the impact of working with the self dual variables instead of the real ones. The result of those investigations, which focus mainly in the context of black hole thermodynamics suggest that the self dual variables could be better behaved than the real ones with respect to the semi classical limit of the theory. It is well known that the computation of the entropy of a spherically symmetric isolated horizon based on the real Ashtekar-Barbero variables leads to a semi-classical result which agrees with the Bekenstein-Hawking area law only up to a fine tuning of the Immirzi parameter $\gamma$. On the other hand, it was shown in \cite{K1, BA1} that the dimension of the Hilbert space of a spherically isolated horizon, which is a function of $\gamma$, can be analytically continued to $\gamma = \pm i$ in a consistent way. Surprisingly, the result matches perfectly the Bekenstein-Hawking entropy without requiring any fine tuning. Although this result has been obtained via an analytic continuation procedure and therefore, there is little control on the underlying self dual quantum theory, it is quite striking. This conclusion was recently generalized to the case of rotating isolated horizon in \cite{RIH}. Aside from this result, others investigations in the context of black holes thermodynamics \cite{MG1, DP1, DP2, HS1, MH, YN1, JBA1, JBA2}, spinfoams models \cite{YN2} and in $(2+1)$-dimensional LQG in \cite{JBA3, JBA4}, were worked out in the same period, all of which suggest that the self dual variables could be more suited than the real ones with respect to the semi classical limit of the theory.

Moreover, from the point of view of the symmetries, it is well known that the real Ashtekar-Barbero connection does not transform as a true space-time connection under the action of the Hamiltonian constraint, contrary to the self dual connection \cite{SM1}. This fact could be responsible for anomalies when going to the quantum theory. In this paper, we will show that working with the self dual variables instead of the real ones indeed allows to preserve the covariance of the system \textit{scalar field coupled to spherically symmetric gravity}, once the holonomy corrections have been introduced. More precisely, it turns out that using the self dual variables allows one to naturally bypass the partial no go result obtained in \cite{S1} with the real variables, and obtain a closed algebra for the modified constraints. Thus, it is possible to obtain an anomaly free algebra for the system \textit{scalar field coupled to spherically symmetric gravity}, which moreover does not exhibit any deformations, reproducing exactly the classical hypersurface deformation algebra of GR\footnote{ It is not necessary that quantum corrections must lead to a deformed notion of general covariance. Indeed, it is known that perturbative higher curvature corrections do not modify the hypersurface deformation algebra \cite{Highercurv, Sasaki}.}. It represents therefore a first step towards the construction of anomaly free LQG model with local physical degrees of freedom, which was up to now out of reach.

From the point of view of the flat spacetime limit, it is well known that the classical hypersurface deformation algebra of GR reduces uniquely to the Poincar\'e algebra \cite{Bojowald}. Since the holonomy corrected algebra of constraint studied in this paper reproduces without deformations the hypersurface deformation algebra of GR, it is natural to conclude that its flat spacetime limit will also reduces to the Poincar\'e algebra. Whether some deformations occurs at the level of the co-algebra in the flat spacetime limit is still an open question \cite{CM}. However, since our interest is to build a quantizable model for studying the gravitational collapse in the context of LQG, the flat spacetime limit of our modified algebra is of no interest for our purposes.

The paper is presented as follows. In section II, we recall the partial no go results obtained in \cite{S1, S2}. In section III, we first introduce the formulation of the spherically symmetric sector of GR written in terms of the self dual Ashtekar variables. This first part relies heavily on the work of Thiemann and Kastrup \cite{T1}. Then we implement holonomy corrections and obtain our modified effective phase space. In section  IV, we study the algebra of the modified constraints and present our main result. Finally, section V is devoted to a discussion of the plausible future direction to quantize this model.

\section{Summary of the partial no go results obtained using the real Ashtekar-Barbero formalism}
\noindent
In this section, we recall some previously established facts and summarize the partial no go results obtained in \cite{S1}.

\subsubsection{The classical framework}

\noindent
Let us consider the phase space of GR written in terms of the real $\SU(2)$ Ashtekar-Barbero variables
\begin{align}
\{ A^i_a, E^b_j\} =  \gamma \delta^i_j \delta^b_a\,, \qquad A^i_a = \Gamma^i_a + \gamma  K^i_a\,, \qquad E^a_i = \sqrt{q} \; e^a_i\,,
\end{align}
where $q$ is the determinant of the induced metric over the spatial slices $q_{ab} = e^i_a e^j_b \delta_{ij}$ and $K^i_a = K_{ab} e^{bi}$ is the extrinsic curvature of these spatial slices. The canonical variables are constrained to satisfy the Gauss constraint $\cG_i$, the vectorial constraint $\cH_a$ and the scalar (Hamiltonian) constraint $\cH$,
\begin{align}
\cG_i = D_a E^a_i\,, \qquad \cH_a = E^b_i  F^i_{ab}\,,  \qquad \cH = \frac{1}{\sqrt{ E} }  \; E_i^a E_j^b  \; ( \; \epsilon^{ij}{}_k F^k_{ab} - 2 ( 1 + \gamma^{2} ) \;K^{[i}_a  K^{j]}_b\; )\,,
\end{align}
where $F^i_{ab} = \partial_a A^i_b - \partial_b A^i_a + (A_a \wedge A_b)^i $ is the curvature of the Ashtekar-Barbero connection.
While the Gauss constraint generates the $\SU(2)$ gauge transformations, the linear combination $D[N^a] = N^a  (\cH_a - \gamma^{-1} A_a . \cG )$  generates the infinitesimal spatial diffeomorphisms  and the scalar constraint $\cH$ generates the infinitesimal diffeomorphisms in the time direction as selected by the initial slicing of the four dimensional manifold $\cM$. Once smeared, those constraints form the following algebra
\begin{align}
& \{ D[M^a_1] , D[M^a_2] \} = D [\cL_{M_1} M^a_2]\,, \\
& \{ H[N], D[M^a]\} = -  H [\cL_M N]\,, \\
& \{ H[N_1], H[N_2] \} =  D [q^{ab}(N_1 \partial_b N_2 - N_2 \partial_b N_1)]\,.
\end{align}
We do not rewrite the Poisson bracket involving only the Gauss constraint, since they will not play a role in what follow.

The spherically symmetric sector of the phase space of GR in terms of the real Ashtekar Barbero variables is given by three pairs of canonically conjugated variables which are constrained to satisfy the three first class constraints $\cG_i$, $\cH_x$ and $\cH$ (there is only one nontrivial component of the Gauss and the diffeomorphism constraint each). Therefore, the number of local physical degrees of freedom is $d = 6 - 2\times 3 = 0$ and we end up with only a global degrees of freedom, the ADM mass, as is expected in vacuum spherical symmetry. The spatial metric is given by
\begin{align}
ds^2 = \frac{(E^{\phi})^2}{|E^x|} dx^2 + |E^x| ( d\theta^2 + \sin^2{\theta} d\phi^2 )\,,
\end{align}
where $x$ is the non compact direction which matches the radial direction at infinity. Once the Gauss constraint $\cG_i$ is solved, we obtain a phase space given by two pairs of canonically conjugated variables denoted (once we set $G=1$) as
\begin{align}
\{ K_x (x), E^x (y)\} = \delta{(x,y)}, \qquad \{ K_{\phi} (x), E^{\phi} (y)\} = \delta{(x,y)}\,.
\end{align}
The remaining constraints are given by
\begin{align}
& D[M] = \int dx \; M(x) \; ( \; \frac{1}{2} (E^x) ' K_x + K'_{\phi} E^{\phi} \; )\,, \\
& H[N] =  \frac{1}{2} \int dx \; N(x) \; \; ( \; \frac{E^{\phi}}{|E^x|^{1/2}} K^2_{\phi} + 2 |E^x|^{1/2} K_{\phi} K_x + \frac{E^{\phi}}{|E^x|^{1/2}} (1 - \Gamma^2_{\phi} ) + 2 \Gamma'_{\phi}  |E^x|^{1/2} \; )\,.
\end{align}

Their algebra is simply the hypersurface deformation algebra adapted to the midisuperspace model, given by
\begin{align}
& \{ D[M_1] , D[M_2] \} = D [\cL_{M_1} M_2] \\
& \{ H[N], D[M]\} = -  H [\cL_M N] \\
& \{ H[N_1], H[N_2] \} = D [q^{xx}(N_1 \partial_x N_2 - N_2 \partial_x N_1)]\,.
\end{align}
Let us now study the holonomy corrected constraints.

\subsubsection{The holonomy corrected constraint algebra for the vacuum case}
\noindent
At the effective level, the polymerization of the connection variables (in this case the extrinsic curvature components), can be implemented as the following transformations
\be
K_x  \rightarrow
f_1(K_x) \qquad \text{and} \qquad K_{\phi} \rightarrow
f _2(K_{\phi})\,.
\ee

However, the holonomy corresponding to $K_x$ is an extended one represented along the edges of  spin-network states and is actually difficult to implement explicitly, requiring suitable non-local functions (see e.g. \cite{MPR} for a negative result in this regard). Therefore, this holonomy correction is usually disregarded in the first attempt. While this might seem to be an oversimplification at first, it can be shown that it is possible to rewrite the constraints in such a manner that the Hamiltonian constraint does not depend on $K_x$. Therefore, the correction $K_x \rightarrow f_1(K_x)$ can be safely ignored for our purposes\footnote{Although the diffeomorphism constraint does still depend on $K_x$, it is usually left unchanged. Even the $K_\phi$ component appearing in $D[N^x]$ is not modified. See \cite{S1} for details on why this is justified.}.

In \cite{Ash1,LM1, LM2, VD1, VD2, C1, CS1, C2, C3, C4}, the function $f_2$ was chosen such that $f_2(K_{\phi} ) = \sin{(\lambda K_{\phi})}/ \lambda$, in analogy with the result of the polymerization procedure that is implemented within LQC. For a formal derivation of this correction, we should first find a regularization formula for the curvature or for the connection in term of the holonomies, such as the Baker Campbell Haussdorf formula used in LQC, and then derive the precise form of the polymerization function $f_2$ from it. In order to avoid this difficulty, we will work with a general function $f_2$ without fixing its expression and allowing for quantization ambiguities. This is the strategy used in \cite{S1}, which ensures that conclusions will be general.

Under this polymerization, the modified algebra of the basic variables is as follow
\begin{align}
\{ K_x (x), E^x (y)\} = \delta{(x,y)}\,, \qquad \{ f_2 (K_{\phi} )(x), E^{\phi} (y)\} = \frac{\text{d} f_2}{\text{d}K_\phi}(x) \delta{(x,y)}\,.
\end{align}
As in \cite{Ash1,LM1, LM2, VD1, VD2, C1, CS1, C2, C3, C4}, we only polymerize the scalar constraint and therefore obtain
\begin{align}
& D[M] = \int dx \;M(x) \;  ( \; \frac{1}{2} (E^x) ' K_x - (K_{\phi} )' E^{\phi} \; )\,, \\
& H[N] =  \frac{1}{2} \int dx \; N(x) \; \; ( \; \frac{E^{\phi}}{|E^x|^{1/2}} f_2 ( K_{\phi} ) + 2 |E^x|^{1/2} g_2 (K_{\phi}) K_x + \frac{E^{\phi}}{|E^x|^{1/2}} (1 - \Gamma^2_{\phi} ) + 2 \Gamma'_{\phi}  |E^x|^{1/2} \; )\,,
\end{align}
where we have introduced a new function $g_2$ in order to be general. One can show \cite{S1} that the functions $f_2$ and $g_2$ are not independent, but are actually related through (which, of course, is true for the classical case as well)
\be
\label{fg}
g_2 = \frac{1}{2} \frac{\partial f_2}{\partial K_{\phi}}\,.
\ee
This property also allows one to Abelianize the constraints, as first proposed in \cite{C2}.

Following \cite{S1}, we compute the algebra of the modified constraints and finally obtains
\begin{align}
& \{ D[M_1] , D[M_2] \} = D [\cL_{M_1} M_2] \,,\\
& \{ H[N], D[M]\} = -  H [\cL_M N]\,, \\
\label{signchange}
& \{ H[N_1], H[N_2] \} = D [ \beta (K_{\phi}) q^{xx}(N_1 \partial_x N_2 - N_2 \partial_x N_1)], \qquad \text{where} \qquad \beta (K_{\phi}) = \frac{1}{2} \frac{\partial^2 f_2}{\partial K^2_{\phi}}\,.
\end{align}

We can see immediately that the algebra of the modified constraints is still closed, allowing for a complete quantization of the spherically symmetric vacuum case. However, the last bracket is modified by the function $\beta$, which depends only on the extrinsic curvature $K_{\phi}$. If the singularity is resolved when $f_2(K_{\phi})$ is maximal, as in LQC, the function $\beta$, given by the second derivative of $f_2$, flips its sign at the `bounce'. This can then be interpreted as an `effective' signature change since then it has the same signature as in the Euclidean case.

Let us now describe the case where a scalar field is coupled to the spherically symmetric vacuum, thus including some local degrees of freedom within the analysis.

\subsubsection{Adding a minimally coupled scalar field}
\noindent
The situation is radically different when we couple a scalar field $\Phi$ to the vacuum model. In this case, the system inherits some local degrees of freedom and the canonical pairs are given by
\begin{align}
\{ K_x (x), E^x (y)\} = \delta{(x,y)}\,, \qquad \{ K_{\phi} (x), E^{\phi} (y)\} = \delta{(x,y)}\,, \qquad \{ \Phi (x) , P_{\Phi}(y) \} = \delta{(x,y)}\,.
\end{align}
The total constraints are written as
\begin{align*}
 D_T[M] & = D_{g}[M] + D_{m}[M]  = \int dx \; M(x) \; ( \; \frac{1}{2} (E^x) ' K_x + K'_{\phi} E^{\phi} \;) + 4 \pi \int dx \; M(x) \;  P_{\Phi} \Phi '\,, \\
H_T[N] & = H_{g}[M] + H_{m}[M] \\
& =  \frac{1}{2} \int dx \; N(x) \; \; ( \; \frac{E^{\phi}}{|E^x|^{1/2}} K^2_{\phi} + 2 |E^x|^{1/2} K_{\phi} K_x + \frac{E^{\phi}}{|E^x|^{1/2}} (1 - \Gamma^2_{\phi} ) + 2 \Gamma'_{\phi}  |E^x|^{1/2} \; ) \\
 & +4 \pi \int dx \; N(x) \; (\;  \frac{P^2_{\Phi}}{2 |E^x|^{1/2} E^{\phi}} + \frac{|E^x|^{3/2}}{2 E^{\phi}} \; \Phi'^{2} + \frac{1}{2} \; |E^x|^{1/2} E^{\phi} \; V(\Phi) \; )\,.
\end{align*}
Proceeding to the same polymerization mentioned above, we obtain the following holonomy corrected constraints
\begin{align*}
D_T[M] & = D_{g}[M] + D_{m}[M] =  \int dx \; M(x) \; ( \; \frac{1}{2} (E^x) ' K_x + K'_{\phi} E^{\phi} \;) + 4 \pi \int dx \; M(x) \;  P_{\Phi} \Phi ' \,\\
\cH_T[N] & = \cH_{g}[M] + \cH_{m}[M] \\
& =  \frac{1}{2} \int dx \; N(x) \; \; ( \; \frac{E^{\phi}}{|E^x|^{1/2}} f_{2}( K_{\phi} ) + 2 |E^x|^{1/2} g_2 (K_{\phi}) K_x + \frac{E^{\phi}}{|E^x|^{1/2}} (1 - \Gamma^2_{\phi} ) + 2 \Gamma'_{\phi}  |E^x|^{1/2} \; ) \\
 & +4 \pi \int dx \; N(x) \; ( \; \frac{P^2_{\Phi}}{2 |E^x|^{1/2} E^{\phi}} + \frac{|E^x|^{3/2}}{2 E^{\phi}} \; \Phi'^{2} + \frac{1}{2} \; |E^x|^{1/2} E^{\phi} \; V(\Phi) \; )\,,
\end{align*}
where the functions $f_2$ and $g_2$ satisfy (\ref{fg}). After a lengthy but straightforward computation, we obtain the algebra of those modified constraints supplemented with holonomy corrections
\begin{align}
\label{11}
 \{ D_T[M_1] , D_T[M_2] \} & = D_T [\cL_{M_1} M_2]\,, \\
 \label{21}
 \{ H_T[N], D_T [M]\} & = -  H_T [\cL_M N]\,, \\
  \label{31}
 \{ H_T[N_1], H_T[N_2] \} & = D_g [ \beta (K_{\phi}) q^{xx}(N_1 \partial_x N_2 - N_2 \partial_x N_1)] + D_m [ q^{xx}(N_1 \partial_x N_2 - N_2 \partial_x N_1)] \\
  \label{41}
& \neq D_T [ \beta (K_{\phi}) q^{xx}(N_1 \partial_x N_2 - N_2 \partial_x \,.N_1)]\,.
\end{align}
While the former two brackets, (\ref{11}) and  (\ref{21}), reproduce the corresponding subalgebra of the  hypersurface deformation algebra, the latter one, (\ref{41}), does not close anymore. Indeed, modifications on the gravitational part occurring through the function $\beta (K_{\phi})$ do not show up in the matter part. Thus we cannot factorize these modifications, and finally obtain the full diffeomorphism constraint $D_T$. One can still try to introduce some modifications through some functions depending on $K_{\phi}$ in the matter part, in such a way that the function $\beta (K_{\phi})$ factorize, but now the bracket between the gravitational and the matter parts is not zero anymore which will prevent the algebra from closing. Note that these results do not depend on the precise form of the correction function $f_2$. Therefore, this is a very general obstruction preventing the matter part of the constraint from closing the algebra. See \cite{S1} for a detailed proof of these statements.

As a consequence, one is forced to conclude that within this context, the holonomy corrected system \textit{scalar field coupled to spherically symmetric gravity} is not covariant. This no go result prevents from quantizing this loop model following the procedure adopted for the vacuum case in \cite{Ash1, LM1, LM2, VD1, VD2,C1, CS1, C2, C3, C4}. It is therefore mandatory, as a first step, to find a way to by pass those partial no go results and find an effective covariant model for this system. In a second step, one should find  a way to get the physical Hilbert space this system. If one succeeds to do so, it would provide a very exciting simplified platform to investigate the fate of a gravitational collapsing scalar quantum shell, its evaporation through Hawking radiation and problems related to those phenomena, such as the information loss paradox.

In this paper, we present a way out for the first step, i.e. the possibility to obtain a covariant holonomy corrected phase space for this system. The second step, i.e. the full quantization of this model, will be discussed at the end but it will require some novel non trivial steps which are still under development and are beyond the scope of this paper.

\section{The spherical symmetric sector of self dual Ashtekar gravity}
\noindent
In this section, we present the spherically symmetric phase space of GR in terms of the self dual Ashtekar variables. The formulation of the spherically symmetric sector was first presented in \cite{T1} where the authors proceeded to the quantization of the spherically symmetric vacuum in the Schr\"odinger representation. We will keep their notations in our presentation. The details concerning the symmetry reduction procedure and the construction of the adapted variables can be found in \cite{MK1, IB1, CT1, CT2, CT3, CT4}.

\subsection{The classical framework}
\noindent
Let us first focus on the pure gravity case. Using the self dual Ashtekar formulation of GR, the Holst action for pure gravity reads (for details, we refer the reader to \cite{T0}).
\begin{align}
S = \frac{1}{\kappa} \int \{ \; \frac{1}{2} \epsilon_{IJKL} e^I \wedge e^J \wedge F^{KL}(A) + \frac{1}{\gamma} e^{I} \wedge e^J \wedge F(A)_{IJ} \; \}\,,
\end{align}
where $\gamma = \pm i$ and the gauge group is $\SL(2, \mathbb{C})$.

The interest of working with self dual variables in the action is that the canonical analysis turns out to be much simpler, and while we end up again with first class constraints, the form of the scalar constraint $\cH$ simplifies drastically. In terms of these constraints, the precedent action can be written as
\begin{align}
S = \frac{1}{\kappa} \int dt \int dx^3\{ \; \Theta_L - ( i \lambda^i \cG_i - i N^a \cD_a + \frac{1}{2} \tilde{N} \cH )\; \}\,,
\end{align}
where the different terms are respectively the canonical variables part, i.e. the Liouville form $\Theta_L$, the Gauss constraint $\cG_i$ enforcing the $\SL(2,\mathbb{C})$ symmetry, the spatial diffeomorphism constraint $\cD_a$ and finally the Hamiltonian constraint $\cH$. The self dual canonical variables are given by
\begin{align}
A^i_a = \Gamma^i_a + i K^i_a\,, \qquad E^a_i = \epsilon^{abc}\epsilon_{ijk} e^j_b e^k_c\,, \qquad \{ A^i_a, E^b_j\} = i\delta^i_j \delta^b_a\,,
\end{align}
and satisfy the first class constraints
\begin{align}
\cG_i  = D_a E^	a_i\,, \qquad \cH_a = \epsilon_{abc} E^{bi} B^c_i \,, \qquad \cH = \frac{1}{2} \;  \epsilon_{abc} \epsilon^{ijk} E^b_i  E^c_j  B^a_k\,,
\end{align}
where the $\tilde{N} = N / (\text{det}\,E)^{-1}$ is the rescaled lapse function. The `magnetic' field variable $B$ has been defined for simplicity and is related to the curvature of the self dual Ashtekar connection as
\be
B^a_i = \frac{1}{2} \delta_{ij} \epsilon^{abc} F^j_{bc} = \frac{1}{2} \delta_{ij} \epsilon^{abc} ( \partial_b A^j_c - \partial_c A^j_b + ( A_b \times A_c )^j )\,.
\ee

Finally, we have also to impose some reality conditions to the canonical variables in order to recover GR:
\begin{align}
 E^i_a E_{b \; i } - \bar{E}^i_a \bar{E}_{b \; i} = 0 \qquad \text{and} \qquad A^i_a - \bar{A}^i_b = 2 \Gamma^i_a\,.
\end{align}
However, as discussed later, there are alternative ways to implement the reality conditions, one of which shall be employed by us.

\subsection{Reduction to spherical symmetry}
\noindent
Let us now select the spherically symmetric sector of this phase space. The procedure for performing this symmetry reduction was presented and discussed in \cite{MK1, IB1, CT1, CT2, CT3, CT4}. We refer the reader to those references for more details. The spatial metric is given by
\begin{align}
ds^2 = \frac{E}{2E^1} dx^2 + E^1 ( d\theta^2 + \sin^2{\theta} d\phi^2 )\,, \qquad \text{where} \qquad E = (E^2)^2 + (E^3)^2\,.
\end{align}

Using this procedure, we end up with the symmetry reduced connection adapted to the spherically symmetric case, as well as its conjugated momentum. They read
 \begin{align}
&  (E^x , E^{\theta}, E^{\phi} ) = ( \; E^1 \sin{\theta} n_x , \; \frac{1}{\sqrt{2}} ( E^2 n_{\theta} + E^3 n_{\phi} ) \sin{\theta} , \; \frac{1}{\sqrt{2}} ( E^2 n_{\phi} - E^3 n_{\theta} ) \; )\,, \\
&  (A_x , A_{\theta}, A_{\phi} ) = ( \; A_1 n_x , \; \frac{1}{\sqrt{2}} ( A_2 n_{\theta} + (A_3 - \sqrt{2}) n_{\phi} )  , \; \frac{1}{\sqrt{2}} ( A_2 n_{\phi} - ( A_3 - \sqrt{2} ) n_{\theta} ) \sin{\theta} \; )\,.
 \end{align}
We can then compute the `magnetic' field $B$
 \be
  (B^x , B^{\theta}, B^{\phi} ) = ( \; B^1 \sin{\theta} n_x , \; \frac{1}{\sqrt{2}} ( B^2 n_{\theta} + B^3 n_{\phi} ) \sin{\theta} , \; \frac{1}{\sqrt{2}} ( B^2 n_{\phi} - B^3 n_{\theta} ) \; )\,,
 \ee
where we have
 \begin{align*}
 & B^1 =  \frac{1}{2} \; ( \;  (A_2)^2 + ( A_3)^2 - 2 )\,,  \\
 & B^2 = A'_3 + A_1 A_2\,, \\
 & B^3 = - A'_2 + A_1 A_3\,.
 \end{align*}
Having obtained our variables, we can now give the expression of the constraints in the reduced symmetry model. The Liouville form reduces to
 \begin{align*}
 \Theta_L & = - i \sin{\theta} \{ \; E^1 \dot{A}_1 + E^2 \dot{A}_2 + E^3 \dot{A}_3 \; \}\,.
 \end{align*}
We deduce that we have three canonically conjugate pair of variables, which define our unconstrained phase space. Those variables are algebraically constrained by the three first class constraints, which therefore lead to $D = 6 - 2 \times 3 = 0$  local degrees of freedom in this system, as expected for the vacuum spherically symmetric sector.

The (nontrivial) constraints are given by:
 \begin{align*}
&  \cG_x
 = \sin{\theta} \; \{ \; (E^1)' - E^2  A_3
 + E^3 A_2  \; \} \,, \\
& \cH_x
= \sin{\theta} \; \{ \; B^2 E^3 - B^3 E^2 \; \} \,, \\
& \cH
= \frac{\sin{\theta}}{2} \{ \; E^2 ( 2 E^1 B^2 + E^2 B^1 ) + E^3 ( 2 E^1 B^3 + E^3 B^1 ) \; \}\,.
 \end{align*}
We can rewrite the diffeomorphism constraint $\cD$ as the following linear combination of the vectorial constrant $\cH_x$ and the Gauss constraint $\cG_x$
\be
\cD = H_x - A_1 \cG_x =  \sin{\theta} \{ A'_3 E^3 + A'_2 E^2 - A_1 (E^1)' \; \}
\ee
This latter will generate the residual diffeomorphisms along the radial direction $x$. Note that the overall factor $\sin{\theta}$ will disappear once integrating over the angular part of the action. Finally, the spherically symmetric version of the reality conditions will not be useful in what follow, thus we refrain from writing them explicitly (see the last section for more details about the reality conditions). Having described our classical symmetry reduced phase space, we can now implement the holonomy corrections.

\subsection{Implementing the holonomy corrections}

\noindent
In this section, we implement the holonomy corrections, which encode the quantum corrections at an effective level inherited from the polymer or loop quantization. However, instead of following exactly what is done in the real formulation \cite{S1}, $K \rightarrow f(K)$, we follow an equivalent procedure where we introduce corrections of the type $B \rightarrow f(B)$. This implies that we end up modifying the curvature functions on using the holonomies (instead of the connection coefficients) in them. Thus, instead of introducing modification functions for the Ashtekar connection components, $A_i$, we therefore choose to modify the dual of the curvature components, $B^i$. Note that this is an equivalent prescription since the dualized curvature components are functions of $A_i$ alone and does not depend on the triad components and, thus, there is a one-to-one correspondence between them. The holonomy corrections, in a very general way, can be implemented by the following transformations
\be
B^1 \rightarrow f_1 (B^1)\,, \qquad B^2 \rightarrow f_2 (B^2)\,,  \qquad B^3 \rightarrow f_3 (B^3)\,.
\ee
However, all of these transformations simultaneously turn out to be too general, and cannot be implemented consistently . Instead, we will mimic what is usually done in the real spherically symmetric model, and introduce only point-wise local holonomy modifications, namely
\be
B^1 \rightarrow f_1 (B^1) \qquad \text{while} \qquad f_2 (B^2) = B^2 \qquad f_3(B^3) = B^3\,.
\ee
As in the real case, we are only modifying the angular part of the curvature, $B^1 \sim F_{23}$. Indeed, since one cannot explicitly compute the holonomy of the $x$-component of the connection $A_1$, one should not modify the curvature component which involve it, i.e. $F_{12}$ and $F_{13}$. One can also understand that point by noticing that implementing local (point-wise) holonomy corrections is equivalent to modifying the angular part of the Ashtekar connection, as in the real variables case. This means introducing general functions of the form $f(A_2^2 + A_3^2)$ in the self dual case, which is equivalent to using a modified version of $B^1$ that represents regularized version of the angular part of the curvature. The dependence of the other curvature components on $A_2$ and $A_3$ are not of the same form and thus modifying those would not be akin to modifying the angular part of the connection alone. This means that if we modify the $A_2$ and $A_3$ components in the other curvature components ($B^2$ and $B^3$), this would imply modifying also the radial components as well, which can only be corrected using non-local functions.

In addition, one can easily check to find that replacing $A_2 \rightarrow h_1(A_2)$ and $A_3\rightarrow h_1(A_3)$ in the $B^2$ and $B^3$ immediately requires that such modification functions have to be the same as the in the classical case from requirements of anomaly freedom of the constraint algebra (See the Appendix for clarifications of this assertion).

With this effective modification, the deformed constraints are therefore given by
\begin{align}
\label{c1}
& \cD = [ A'_3 E^3 + A'_2 E^2 - A_1 (E^1)' \; ]\,, \\
\label{c3}
&  \cH = \frac{1}{2}  \; [ \; E^2 ( 2 E^1 B^2 + E^2 f_1(B^1) ) + E^3 ( 2 E^1 B^3 + E^3 g_1 (B^1) ) \; ]\,,
\end{align}
where we have introduced a new function $g_1$ to remain general. We will see that, in order to close the algebra, we must require $f_1 = g_1$.
Note also that, once again, we have modified only the scalar constraint, following what was done in \cite{S1}\footnote{In our case, this seems natural as we are only modifying the curvature components and not the connection components directly. Indeed, the magnetic field component $B^1$ doesn't enter in the expression of the diffeomorphism constraint.}.
We are now ready to compute the algebra of the modified constraint and investigate the covariance of this new holonomy-corrected system.

\section{Investigating the covariance of the model in the self dual context}
\noindent
In this section we present the computation of the algebra of the modified constraints. We first demonstrate our results for the vacuum case, and then extend it to the system composed by the scalar field minimally coupled to spherically symmetric gravity.

\subsection{The vacuum model}

\noindent
The Hamiltonian constraint, for this case, reads
\bea
H[N]&=&\left(\frac{1}{2}\right)\int \text{d}x N(x) \left\{E^2(x)\left(2B^2(x)E^1(x) + B^1(x)E^2(x)\right)\right. \nn\\
& &\,\,\,\qquad\left.+ E^3(x)\left(2B^3(x)E^1(x) + B^1(x)E^3(x)\right)\right\}\,.
\eea
Due to regularization, we need to introduce modification functions for  implementing the holonomy corrections. As explained earlier, we only want to introduce the local holonomy corrections by replacing the above constraint by
\bea
H[N]=\left(\frac{1}{2}\right)\int \text{d}x N  \left\{E^2\left(2B^2E^1 + f_1\left(B^1\right)E^2\right) + E^3\left(2B^3E^1 + g_1\left(B^1\right)E^3\right)\right\}\,,
\eea
where we have introduced the two different modification functions $f_1$ and $g_1$ and have suppressed the argument of each of the phase space variables on the radial coordinate. The other constraints, the diffeomorphism one and the Gauss one, remain unmodified from the classical ones.  Our first goal is to calculate the $\left[H[N], H[M]\right]$ bracket with such correction functions. We look at the brackets of the Hamiltonian constraint with itself and with the diffeomorphism constraint. The other bracket involving the Hamiltonian constraint with the Gauss constraint remains obviously unmodified.

\subsubsection{$[H,H]$ bracket}

\noindent
We start with the calculation of the $[H,H]$ bracket, since this is the only one which gets deformed in the real Ashtekar-Barbero case by a modification of the structure functions appearing on the right hand side of the expression. In such calculations, it is useful to remember that we have a non zero bracket between two conjugate variables only when one of them have a spatial derivative on it. (For instance, note that $B^1$ is independent of spatial derivatives whereas $B^2, B^3$ are not.) Thus, the contribution to this bracket should come from the commutator between the first and third term, the second and third term and the first and fourth term in the Hamiltonian constraint.

The bracket between the first and third term is
\bea\label{TermA}
& &\int \text{d}x \text{d}y M(x)N(y) E^1(x)E^1(y)\left[E^2(x)B^2(x), E^3(y)B^3(y)\right] - (x \leftrightarrow y)\nn\\
&=& i\int \text{d}x \text{d}y M(x)N(y) E^1(x)E^1(y)\left\{E^2(x)B^3(y) \frac{\text{d}}{\text{d}x}\left[\delta(x,y)\right] + B^2(x)E^3(y) \frac{\text{d}}{\text{d}y}\left[\delta(x,y)\right]\right\} - (x \leftrightarrow y) \nn\\
&=& i\int \text{d}x \left(M'(x)N(x)-N'(x)M(x)\right) \left\{\left(E^1(x)\right)^2 E^2(x) B^3(x) - \left(E^1(x)\right)^2 E^3(x) B^2(x)\right\}\,.
\eea
The above bracket does not involve any of the modification functions and is exactly what it would have been for the classical case. Turning our attention to the bracket between the first and fourth term, we find the appearance of the holonomy corrections
\bea\label{TermB}
& &\int \text{d}x \text{d}y M(x)N(y) E^1(x)E^2(x)f_1\left(B^1(y)\right)\left[B^2(x), \left(E^3(y)\right)^2\right] - (x \leftrightarrow y)\nn\\
&=&\int \text{d}x \text{d}y M(x)N(y) E^1(x)E^2(x)f_1\left(B^1(y)\right)E^3(y) \frac{\text{d}}{\text{d}x}\left[\delta(x,y)\right] - (x \leftrightarrow y)\nn\\
&=& i\int \text{d}x \left(M'(x)N(x)-N'(x)M(x)\right) E^1(x)E^2(x)E^3(x)f_1\left(B^1(y)\right)\,.
\eea
Proceeding similarly, we find that the bracket between the second and third terms gives
\bea\label{TermC}
-i\int \text{d}x \left(M'(x)N(x)-N'(x)M(x)\right) E^1(x)E^2(x)E^3(x)g_1\left(B^1(y)\right)\,.
\eea

We know that $\left[H[N], H[M]\right]$ must close into one of the other first class constraints. In the classical case we have
\bea
\left[H[N], H[M]\right] = i\int \text{d}x \left(M(x)N'(x)-N(x)M'(x)\right) \left(E^1(x)\right)^2 \left\{B^2(x)E^3(x) - B^3(x)E^2(x)\right\}\,,
\eea
where the vector constraint is given by $H^x[N^x] = \int \text{d}x N^x(x)\left\{B^2(x)E^3(x) - B^3(x)E^2(x)\right\}$.
In order to have a similar closure of the $\left[H[N], H[M]\right]$ commutator, it is immediately obvious that we require $f_1\left(B^1\right) = g_1\left(B^1\right)$ for (\ref{TermB}) to cancel (\ref{TermC}), just as in the classical case. Thus we find that the the commutator of Hamiltonian constraints give us the exact same result as we have in the classical case, even in the presence of holonomy modifications!

There are a few observations to make from this rather astonishing result. Firstly, from a mathematical point of view, the reason for this can be explained as follows. The modification to the structure functions in the real variables case is primarily due the presence of second (spatial) derivatives of the triad components appearing in the Hamiltonian constraint \cite{InformationLoss}. Those terms appear from the presence of the spin connection term, which does not appear in the self-dual case. The coefficient of the spin connection term is $(1+\gamma^2)$ and that goes to zero when $\gamma=i$. (This also tells us that this result is a rather special case and would not be valid for a general imaginary Immirzi parameter.) The other thing to point out is that we should really look at the hypersurface deformation algebra which really has the $[H,H]$ bracket closing into a diffeomorphism constraint. We can easily rewrite the vector constraint as a combination of the diffeomorphism constraint and the Gauss constraint. We then need to make sure that the bracket of the Hamiltonian constraint closes with both these other first class constraints. The one with the Gauss constraint remains obviously unmodified whereas the one with the diffeomorphism constraint is shown in the next section.

\subsubsection{$[D,H]$ bracket}

\noindent
Rewriting the diffeomorphism constraint
\bea
D[N^x]=\int \text{d}x N^x [-A_1(E^1)^\prime + A_2^\prime E^2 + A_3^\prime E^3]\,,
\eea
we want to evaluate the bracket $\left[D[N^x], H[N]\right]$. Instead of explicitly showing the full calculation involving all the terms, let us only focus on the bracket of the second term from the Hamiltonian constraint with the diffeomorphism constraint. It is enough to do so in this case since a particular term of the Hamiltonian constraint must reproduce that specific term from the bracket with the diffeomorphism constraint:
\bea
&&\left[D[N^x], -\frac{1}{2}\int \text{d}y N f_1(B^1)(E^2)^2\right]\nn\\
&=& \frac{1}{2}\int\text{d}x\text{d}y N^x(x)N(y) \left[-A_1(x)(E^1)^\prime(x)+ A_2(x)^\prime E^2(x)+A_3^\prime(x) E^3(x), f_1(B^1(y))(E^2)^2(y)\right]\nn\\
&=& -i\int\text{d}x N^{x\prime} N f_1(B^1) (E^2)^2 -i\text{d}x N^xN f_1(B^1) (E^2)^\prime E^2\nn\\
& &\,\,\, -\frac{i}{2}\int \text{d}x N^xN A_2^\prime \frac{\partial f_1}{\partial A_2}(E^2)^2 -\frac{i}{2}\int \text{d}x N^xN A_3^\prime \frac{\partial f_1}{\partial A_3} (E^2)^2\nn\\
&=& -\frac{i}{2}\int \text{d}x \left(N^{x\prime} N - N^x N'\right) (E^2)^2 f_1(B^1) -\frac{i}{2}\int \text{d}x N^{x\prime} N (E^2)^2 f_1(B^1) -\frac{i}{2}\int \text{d}x  N^x N'(E^2)^2 f_1(B^1)\nn\\
& &\,\,\, -\frac{i}{2}\int \text{d}x N^x N (E^2)^2 [f_1(B^1)]^\prime -i\int \text{d}x N^xN f_1(B^1) (E^2)^\prime E^2\nn\\
&=& -\frac{i}{2}\int \text{d}x \left((N^x)^\prime N - N^x N'\right) (E^2)^2 f_1(B^1) -\frac{i}{2}\int \text{d}x (N^xN)^\prime (E^2)^2 f_1(B^1)\nn\\
& &\,\,\, -\frac{i}{2}\int \text{d}x N^xN (E^2)^2 [f_1(B^1)]^\prime -\frac{i}{2}\int \text{d}x N^xN f_1(B^1)\left((E^2)^2\right)\prime\nn\\
&=& -\frac{i}{2}\int \text{d}x \left((N^x)^\prime N - N^x N'\right) (E^2)^2 f_1(B^1) + \text{total derivative}\,.
\eea
Similarly, the bracket of all the other terms of the Hamiltonian constraint with the diffeomorphism constraint reproduces the whole Hamiltonian constraint.

Thus, we have shown that starting from the holonomy corrected constraints
\begin{align}
\label{ca1}
& \cD = [ A'_3 E^3 + A'_2 E^2 - A_1 (E^1)' \; ]\,, \\
\label{ca3}
&  \cH = \frac{1}{2}  \; [ \; E^2 ( 2 E^1 B^2 + E^2 f_1(B^1) ) + E^3 ( 2 E^1 B^3 + E^3 g_1 (B^1) ) \; ]\,,
\end{align}
their algebra is given by
\begin{align}
\label{1sd}
 \{ D[M^x_1] , D[M^x_2] \} & = D [\cL_{M^x_1} M^x_2]\,, \\
 \label{2sd}
 \{ H[N], D [M^x]\} & = -  H [\cL_{M^x} N]\,, \\
  \label{3sd}
 \{ H[N_1], H[N_2] \} & =   H^x [  (E^1(x))^2(N_1 \partial_x N_2 - N_2 \partial_x N_1)]\,,
\end{align}
reproducing without any modifications the classical algebra of constraints. However, to reproduce the hypersurface deformation algebra of GR, we need to rewrite the vector constraint in terms of the diffeomorphism constraint and the Gauss constraint as well as restore the rescaling of the lapse functions in terms of the determinant of the spatial metric. Looking at the RHS of (\ref{3sd}),
\bea
& &\int \text{d}x \left(E^1(x)\right)^2
\left(N_1\partial_x N_2 -N_2 \partial_x N_1\right) \cH^x\nn\\
&=& \int \text{d}x \left(E^1(x)\right)^2 \left(\frac{\tilde{N}_1}{(\text{det}E)}\partial_x\left( \frac{\tilde{N}_2}{(\text{det}E)}\right) - \frac{\tilde{N}_2}{(\text{det}E)} \partial_x\left(\frac{\tilde{N}_1}{(\text{det}E)}\right)\right) \left(\cD + A_1\cG^x\right)\nn\\
&=& \int \text{d}x \,q^{xx} \left(\tilde{N}_1\partial_x \tilde{N}_2 -\tilde{N}_2 \partial_x \tilde{N}_1\right) \left(\cD + A_1\cG^x\right)\,,
\eea
it is clear that we can recover the usual form of the classical hypersurface deformation algebra, once we solve for the Gauss constraint $\cG^x\approx 0$. (We have used the expression for $q^{xx}$ in the last line above.)

More precisely, the modifications function $\beta(K_{\phi})$ that shows up in the Poisson bracket between two Hamiltonian constraints (\ref{signchange}) in the real formulation has disappeared in the self dual formulation. Therefore, the question of the interpretation of this modification simply drops out in the context of the self dual formulation.

\subsection{Adding a minimally coupled scalar field}
\noindent
Let us now couple a scalar field to the spherically symmetric vacuum. The preceding holonomy corrected constraints for the vacuum case are now supplemented by terms arising from the matter part. The matter part of the constraint comes in turn from the corresponding ones for a minimally coupled scalar to a spherically symmetric space-time. The Hamiltonian constraint for a scalar field is given by

\bea
H_{\text{scalar}}=\int \text{d}^3x N \left\{  \frac{P_\Phi^2}{2\sqrt{q}} - \frac{1}{2}\sqrt{q}q^{xx}\Phi'^2 +\sqrt{q} V(\Phi)  \right\}\,,
\eea
whereas the diffeomorphism constraint in this case is
\bea
D_{\text{scalar}}=\int \text{d}^3x N^x \Phi' p_\Phi \,.
\eea
We have one more canonical pair in the new phase space given by $\{ P_\Phi(x), \Phi(y) \} = \delta(x,y)/4\pi$.
Thus the full constraint is as follow
\begin{align*}
D_T[M^x] & = D_{\text{grav}}[M^x] + D_{\text{scalar}}[M^x] =  \int dx \; M(x) \; ( \; A'_3 E^3 + A'_2 E^2 - A_1 (E^1)'  \;) + 4 \pi \int dx \; M(x) \;  P_{\Phi} \Phi ' \,,\\
H_T[N] & = H_{\text{grav}}[N] + H_{\text{scalar}}[N] \\
& =  \frac{1}{2} \int dx \; N(x) \; \; ( \; E^2 ( 2 E^1 B^2 + E^2 f_1(B^1) ) + E^3 ( 2 E^1 B^3 + E^3 f_1 (B^1) ) \; ) \\
 & +2 \pi \int dx \; N(x) \; \left( \; P^2_{\Phi} + (E^1)^2 \; \Phi'^{2} + E^1((E^2)^2+(E^3)^2)  \; V(\Phi) \; \right)\,.
\end{align*}

It is easy to observe that there shall be no holonomy modification functions appearing in the matter sector since they do not have any dependence on the connection components but rather on only triad variables. Thus the commutator of the scalar Hamiltonian would close into the scalar diffeomorphism, as in the classical case. However, deviations from the classical form of the hypersurface deformation algebra can still take place only if there are cross terms between the gravity and matter sectors, i.e. if $\left[H_{\text{grav}}[N], H_{\text{scalar}}[M]\right]\neq 0$. In the classical case, such cross terms were zero and thus the bracket closed into the vector constraint. The holonomy corrections introduced by us have an argument of $B^1$, and do not depend on the other curvature components. However, $B^1$ does not have any spatial derivatives on the connection components, which has no non-zero contribution to the above bracket. Since this is the only modification introduced by us, and it does not contribute to the bracket, we can easily conclude $\left[H_{\text{grav}}[N], H_{\text{scalar}}[M]\right]= 0$. Thus the full Hamiltonian constraint commutator still closes into constraints without any deformation of the structure function as in the classical case.

All the other brackets between the full Hamiltonian constraint and the full diffeomorphism constraint, and between the full Hamiltonian constraint and the full Gauss constraint remains the same as in the classical case. We can see this without explicit calculation by just noticing that all the cross terms between the gravitational and the matter parts of the constraints vanish, i.e. $\left[H_{\text{grav}}[N], D_{\text{scalar}}[N^x]\right]=0$ and $\left[H_{\text{grav}}[N], \cG_{\text{scalar}}[\lambda]\right]= 0$. Thus we now have a model with local degrees of freedom which does have holonomy corrections but the underlying covariance of the system is the same as the classical case and is given by
\begin{align}
\label{1}
 \{ D_T[M_1] , D_T[M_2] \} & = D_T [\cL_{M_1} M_2] \,, \\
 \label{2}
 \{ H_T[N], D_T [M]\} & = -  H_T [\cL_M N] \,,\\
  \label{3}
 \{ H_T[N_1], H_T[N_2] \} & =  D_T [ q^{xx}(N_1 \partial_x N_2 - N_2 \partial_x \, N_1)]\,,
\end{align}

where we have written $H_T = H_{\text{grav}} + H_{\text{scalar}}$ and $D_T = D_{\text{grav}} + D_{\text{scalar}}$.

\section{Discussion}
\noindent
Let us now discuss the results obtained in this paper. We have shown that the partial no go results obtained in \cite{S1} within the context of spherically symmetry reduced loop model can be overcome by working with the self dual variables instead of the real ones. We comment briefly on the status of the self dual variables and then discuss the possibility to quantize this model.

\subsubsection{Undeformed covariance with self-dual variables}
\noindent
From the point of view of the self dual variables, our result reinforces the idea that the self dual variables are more natural in the context of the loop quantum theory of black holes. Indeed, as pointed in the introduction, it has been shown recently that the self dual variables reproduce in a more satisfying way (without any fine tuning) the expected semi-classical results in the context of black holes thermodynamics, such as the Bekenstein-Hawking area law for the entropy of spherically and rotating isolated horizons \cite{K1, BA1, RIH}, or the thermal character of the partition function for the horizon \cite{MG1, DP1, DP2}. In this paper, we have shown that, additionally, the use of the self dual variables allows to define a quantizable model for the gravitational collapse, which does not suffer from the drawbacks of \cite{M1} and by pass naturally the no go result found in \cite{S1}.

While the model we studied in this work is an effective quantum one, where the holonomy corrections are partially implemented (recall that we only modified the angular part of the curvature), it represents the first holonomy corrected model of spherically symmetric gravity coupled to a scalar field which is fully covariant. It seems that one is forced to use the self dual variables in order to define a quantizable holonomy corrected model for this system due to the no-go theorem of \cite{S1}.

Even more remarkably, we have shown that the holonomy corrected algebra has the exact same form as the classical hypersurface deformation algebra, which is an unexpected outcome. Indeed, it is commonly believed that classical diffeomorphism symmetry of general relativistic spacetimes will be deformed at the quantum level. While it is the case when one uses the real Ashtekar-Barbero variables resulting in the so called signature change phenomenon \cite{S1}, it turns out that no symmetry deformation occurs in the self dual case, at least in the spherically symmetric sector. Whether this conclusion extends to other sectors where the signature change phenomenon also occurs, needs to be worked out in the future. If this conclusion holds for all such models, then one should interpret signature change as an artificial characteristic of using the real Ashtekar-Barbero variables.

Moreover, the fact that we have obtained an algebra of the modified constraints which is exactly the same than the hypersurface deformation algebra, as arising for ordinary GR, could seem problematic at first. Indeed, the well known theorem obtained by Hojman, Kuchar and Teitelboim \cite{HKT, Kukar} states that starting from the hypersurface deformation algebra, one can uniquely derive the Einstein-Hilbert action, up to a  cosmological constant, when the constraints contain no higher than second derivatives of the metric. However, that result has been derived for full $(3+1)$-gravity and we have considerable more freedom in our case since we are working in a symmetry reduced model. (Once again, this is related to the fact that we have only one non-zero component of the spatial diffeomorphism constraint in our case.) Therefore, the Hojman-Kuchar-Teitelboim theorem, in its original form, doesn't apply to our symmetry reduced system.

Finally, from the point of view of the anomaly free algebra, this result opens up a very promising path towards constructing quantum theories of inhomogeneous midisuperspace models with local physical degrees of freedom, which have been out of reach up till now. While our results point towards the need of using self dual variables for such non trivial models\footnote{Indeed, we have demonstrated a similar result for cosmological scalar perturbations in the self dual context where we end up with an undeformed algebra, as opposed to the system with real-valued variables \cite{SDLQCP}.}, a definitive conclusion is not yet possible and one has to investigate some other reduced loop model in order to obtain a more robust generic result along these lines. The next interesting reduced models to investigate would be polarized Gowdy models for which partial no go theorem have already been proven in \cite{S2} using the real Ashtekar-Barbero formulation. We plan to address this question in a future work.

While very encouraging, the anomaly free algebra presented in this work will be useful only if one is either able to extract a concrete effective theory from it, by picking up an explicit form of the holonomy correction $f_1$, or, even better if one is able to quantize this model based on the self dual variables. Let us now comment on this point.

\subsubsection{On the quantization of this self-dual model}
\noindent
There are two main outstanding technical complications that are encountered within the quantization procedure of this model:
\begin{enumerate}
\item one has to derive the explicit expression of the holonomy correction function $f_1 (B^1)$;
\item one has to find a way to implement the reality conditions inherent to the self dual formulation.
\end{enumerate}
While the first point represents the most important difficulty, the second one can be overcome more easily in this spherically symmetry reduced case. Let us comment on this latter point first. While the imposition of the reality conditions at the quantum level for the full theory is a highly non trivial problem, which remains open since the very advent of the self dual variables, the situation is quite different in symmetry reduced models. Indeed, if one knows what are the quantum observables of the system studied, which is a non trivial question for a diffeomorphism invariant system, one can simply require those observables to be self adjoint with respect to the scalar product on the physical Hilbert space of the quantum theory. This was precisely the strategy used in \cite{T1} where the spherically symmetric self-dual Ashtekar gravity was quantized in the Schr\"odinger representation. Therefore, it seems reasonable to infer that the problem generally associated to the reality conditions will not be difficult to be solved within our symmetry-reduced model either.

For the first obstruction, the situation is more subtle. Since we are now working in the self-dual formulation of gravity, the gauge group is non compact and given by $\SL(2,\mathbb{C})$. At first sight, it could seem hopeless to expect a resolution of the singularity because the holonomy corrected functions can be unbounded contrary to the $\SU(2)$ case. However, it has been shown in \cite{JBA5} that such preconceptions  can be misleading. In this work, a proposal for defining a self dual model of LQC through an analytic continuation procedure was introduced. It was possible to exhibit a bounded holonomy correction function involving hyperbolic functions, which still preserved the bouncing scenario and led to the right semiclassical limit. Although the precise expression of this function is quite complicated and a full quantization of the model becomes rather intractable, it shows that we have not to work necessarily with almost-periodic trigonometric functions in order to obtain the bouncing scenario at the effective level. This issue can be better understood as follows. Because of the non compactness of the gauge group, it is now possible to work either with null, elliptic or hyperbolic elements of the group when computing the holonomy corrections. Since the graphs, on which the quantum states are supposed to be defined, live on a space-like hypersurface due to the initial slicing of the manifold, the holonomy of the connection associated to each edge will be given by an hyperbolic element of $\SL(2,\mathbb{C})$. It is then straightforward to see that we cannot work in the fundamental representation if we want to obtain singularity resolution  since the curvature will be replaced by an unbounded function. Instead, we need to  investigate the higher dimensional representations of the group in order to obtain holonomy corrections which remains bounded (albeit this comes at the expense of working with more complicated looking functions).
Implementing this program is currently under investigation with promising initial results. \\

 \section*{Acknowledgements}
We are indebted to Martin Bojowald for a careful reading of an earlier version of the draft and for many helpful suggestions on it. We would also like to thank Miguel Campiglia, Rodolfo Gambini, Jorge Pullin and Javier Olmedo for useful comments, as well as Julien Grain, Giovanni Amelino-Camelia and Michele Ronco for critical remarks. This work has been supported by the Shanghai Municipality, through the grant No. KBH1512299, and by Fudan University, through the grant No. JJH1512105.

\section*{Appendix}
\noindent
Let us suppose that we do not consider local holonomy corrections alone and introduce modification fucntions in $A_2$ and $A_3$ wherever we see them in $B^1, B^2$ and $B^3$. The correction functions would then take the form
\bea
B^1 &=& f_1(A_2^2 + A_3^2)\\
B^2 &=& \frac{{\rm d} g_1(A_3)}{{\rm d} A_3}A_3^\prime + A_1h_1(A_2)\\
B^3 &=& \frac{{\rm d} g_2(A_2)}{{\rm d} A_2}A_2^\prime + A_1h_2(A_3)\,.
\eea
The correction functions $g_1$ and $g_2$ an be immediately ruled out by looking at Eqns (\ref{TermB}) and (\ref{TermC}). If we have $g_1$ and $g_2$ different from the classical case, then these two terms cannot cancel out any more (which we require for the closure of the brackets).

However, the modification functions $h_1$ and $h_2$ do not cause any problems as far as the closing of the brackets are concerned. But if we look at Eqn (\ref{TermA}), we shall find that the new term left over is not going to be proportional to the diffeomorphism (or vector) constraint any longer. The new version of Eqn (\ref{TermA}) shall be given by
\bea
i\int \text{d}x \left(M'(x)N(x)-N'(x)M(x)\right) \left\{\left(E^1(x)\right)^2 E^2(x) B^3(x) - \left(E^1(x)\right)^2 E^3(x) B^2(x)\right\}\,,
\eea
with the modification coming from the fact that $B^1$ and $B^2$ are now modified by the presence of $h_1$ and $h_2$. One might think that this is okay if we redefine our vector constraint to have these modifications in built in it. However, in that case, the diffeomorphism constraint shall also get modified by the presence of these functions $h_1$ and $h_2$ (remember that the diffeomorphism constraint is just a linear combination of the vector and Gauss constraints). It is then easy to convince one self that the $\left\{D,D\right\}$ bracket does close with any modification functions in it (and this is partly the reason why we do not modify the diffeomorphism constraint). Thus, from an independent perspective, it is easy to see that we cannot close the bracket with local correction functions beyond what we have introduced in this paper for $B^1 \rightarrow f(B^1)$.


\end{document}